\documentclass[preprint,nofootinbib]{revtex4}

\usepackage{graphicx}
\usepackage{multirow}
\usepackage{xcolor}

\begin{document}

\title{Joint analysis of EDGES $21$-cm line observations with standard candles and rulers in $\Lambda$CDM and non-adiabatic gCg models}

\author{C. Pigozzo$^{1}$, S. Carneiro$^{1}$ and J. C. Fabris$^{2,3}$}

\affiliation{$^1$Instituto de F\'{\i}sica, Universidade Federal da Bahia, 40210-340, Salvador, Bahia, Brazil\\$^2$N\'ucleo Cosmo-ufes \& Departamento de F\'{\i}sica - UFES, 29075-910, Vit\'oria, ES, Brazil\\$^3$National Research Nuclear University MEPhI, Moscow, Russia}

\date{\today}

\begin{abstract}
A decomposed generalised Chaplygin gas (gCg) with energy flux from dark energy to dark matter, represented by a negative value for the gas parameter $\alpha$,  is shown to alleviate the tension between EDGES data and the cosmological standard model.  Using EDGES data and employing a Bayesian statistical analysis, the agreement with the standard model is only marginal. However, if $\alpha$ is negative enough the gCg fits remarkably well the data, even in combination with SNe Ia datasets.  On the other hand, when the CMB and BAO acoustic scales are included the preferred value for $\alpha$ is near zero, implying that a small deviation from $\Lambda$CDM is predicted.
\end{abstract}

\maketitle


EDGES observations of the hydrogen $21$-cm line transition are in tension with the standard cosmological model \cite{Nature}. In spite of some discussion about the foreground subtraction \cite{critica1,critica2}, this has lead to many proposals of explanations based on alternative cosmologies. The tension is alleviated for example in models with energy flux from dark energy to dark matter, because, for the same present amount of matter, in these models the Hubble function is attenuated at intermediate redshifts (see e.g. \cite{chineses}). In this short note we present a joint analysis of EDGES data combined with standard candles and rulers, in the broader context of a non-adiabatic generalised Chaplygin gas, obtaining a good concordance for negative values of the Chaplygin parameter, that is, an energy flux from dark energy to dark matter\footnote{For an analysis with only positive values of the Chaplygin parameter and with a modified gCg, see \cite{valentina}. In the modified gCg negative values are also allowed. As the best-fits constrain the additional parameter $B$ to approximately zero, one could conclude that the modified gCg from \cite{valentina} is equivalent to the gCg. This would be true if $B$ were set to zero before the data fitting. This is not the case, because each parameter in Table III of \cite{valentina} is determined by marginalising the other parameters, not fixing them.}.


The $21$-cm line brightness temperature relative to the CMB background temperature $T_{\text{CMB}}$ is given by \cite{Nature,chineses}
\begin{equation} \label{1}
T_{21}(z) = \frac{T_{\text{S}}(z) - T_{\text{CMB}}(z)}{1+z}\, \tau_{\nu_0}(z),
\end{equation}
with
\begin{equation} \label{2}
\tau_{\nu_0}(z) \approx 0.053\, x_{\text{HI}}\, \Omega_{b0}\, h \left[ \frac{T_{\text{CMB}}(z)}{T_{\text{S}}(z)} \right] \frac{(1+z)^2}{E(z)},
\end{equation}
where $E(z) = H(z)/H_0$ is the adimensional Hubble function, and $H_0 = 100\, h$ km/s.Mpc.  At $z = 17.2$ it was measured by EDGES as $\hat{T}_{21} = -500^{+200}_{-500}$ mK ($3\sigma$), in tension with the standard model prediction $T_{21} \approx -209$ mK. In the above formulae, $T_{\text{S}}$ is the $21$-cm spin temperature, assumed to be equal to the gas temperature at this redshift, $T_{\text{S}}  \approx 7.3$ K. The fraction of neutral hydrogen is assumed to be $x_{\text{HI}} = 1$. 
Our goal is to evidence that EDGES data favours a flux of energy from Dark Energy to Dark Matter, that here will be modeled by a decomposed gCg with clear preference to a negative interaction parameter $\alpha$. This main result does not change if we relax some approximations adopted. For example, using the exact expressions for the spin temperature given in \cite{ApJ} it was verified, for a wide range of the gCg parameter, that $T_{\text S} (z = 17.2)$ differs by less than $1\%$ from the $\Lambda$CDM value \cite{pacheco}.

Once the errors on the 21 cm-line measurement are asymmetrical, the log-likelihood adopted was a variable Gaussian \cite{asymmetric} written as
\begin{equation}
	\ln L = -\frac{1}{2}\frac{[\hat{T}_{21}-T_{21}(z=17.2)]^2}{V+[T_{21}(z=17.2)-\hat{T}_{21}]V'} ,
\end{equation}
where $\hat{T}_{21}$ is the measured temperature  and
\begin{equation}
	V = |\sigma_-|\sigma_+ \quad \rm{and}  \quad  V' = \sigma_+-|\sigma_-|
\end{equation}
($\sigma_-$/$\sigma_+$ are the positive/negative errors on $\hat{T}_{21}$). When $|\sigma_-|=\sigma_+$ it recovers the usual symmetric Gaussian log-likelihood. The asymmetric likelihood is shown in Fig. 1. The alternative use of other possible asymmetric likelihoods presented in \cite{asymmetric} does not change our results substantially.


\begin{figure*}
 \begin{center}
 \includegraphics[width=0.45\textwidth]{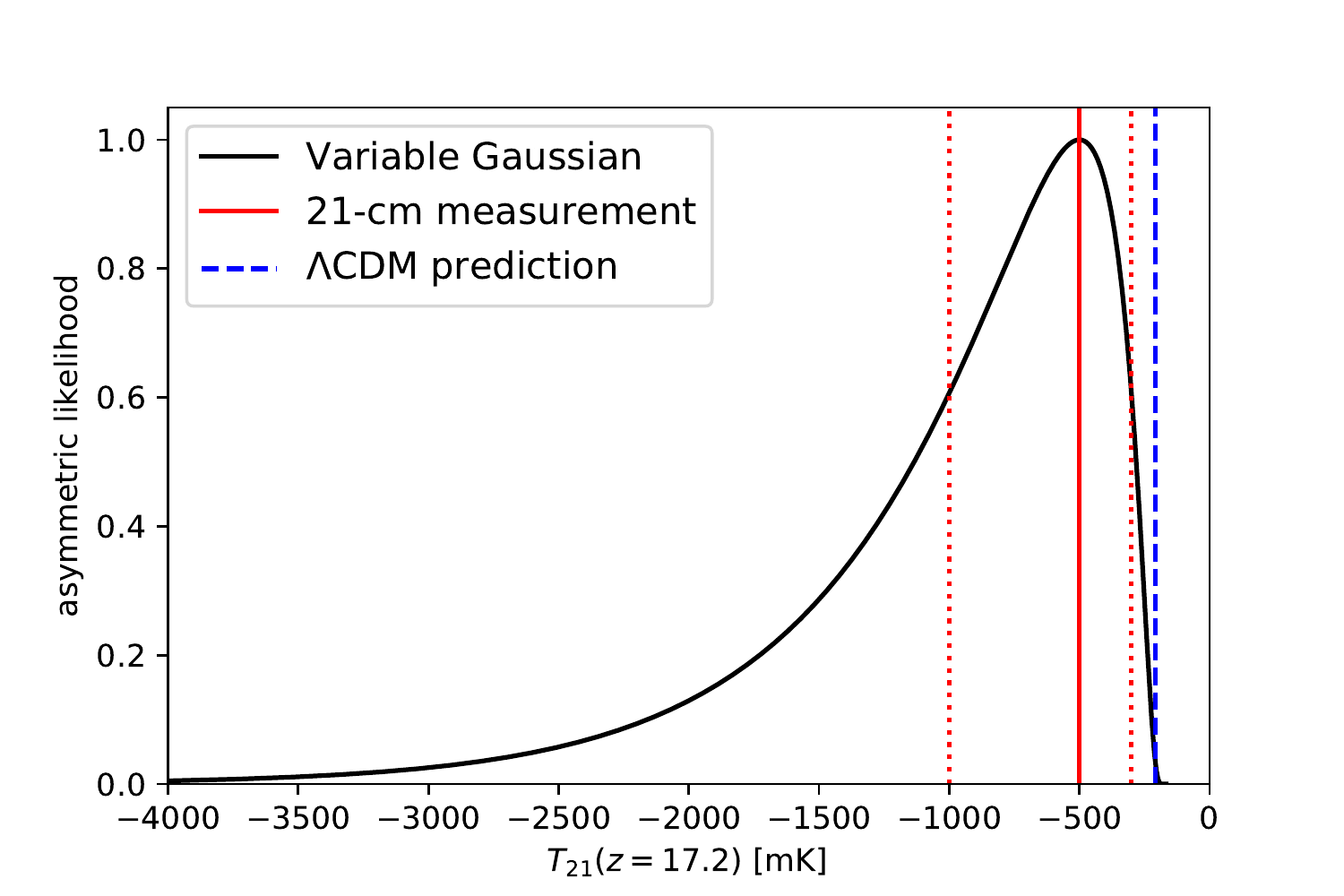} 
  \end{center}
  \vspace{-0.3in}
  \caption{Variable Guassian likelihood describing asymmetric errors for the 21 cm-line measurement.}
 \label{asymmetric}
 \end{figure*}

We will include two standard rulers in our analysis of EDGES results. The first is given by the position of the first peak in the CMB spectrum of anisotropies, more precisely the CMB acoustic scale
\begin{equation}
{\theta_*} =  \frac{r_{ls}}{D_A},
\end{equation}
where $D_A$ is the comoving angular diameter distance to the last scattering surface,
\begin{equation}
D_A = \int_0^{z_{ls}} \frac{c}{H(z)},
\end{equation}
and $r_{ls} = r_s(z_{ls})$ is the acoustic horizon
\begin{equation}
r_s(z) = \int_z^\infty \frac{c_s}{H(z')} dz'
\end{equation}
at the last scattering. The sound speed is given by
\begin{equation}
\frac{c_s}{c} = \left[ 3 + \frac{9\Omega_{b0}}{4\Omega_{\gamma 0}} (1+z)^{-1} \right]^{-1/2},
\end{equation}
where $\Omega_{b0}$ and $\Omega_{\gamma 0} = 2.47 \times 10^{-5}\, h^{-2}$ are the density parameters of baryons and photons, respectively. 
The observed value of the angular scale is $100\theta_* = 1.04109 \pm 0.00030$ \cite{Planck}. The second ruler, related to BAO's observations, is the acoustic horizon $r_d = r_s (z_d)$ at the drag epoch, determined by Verde {\it et al.} as $r_d^h \equiv r_d h = (101.2  \pm 2.3)$ Mpc \cite{Verde}. Both $z_{ls}$ and $z_d$ were evaluated from recombination fitting formulae \cite{fittingformulae}. In order to verify their possible dependence on the models under test, we have derived their values from the Planck chains for several dark energy models, and no appreciable difference was found. In fact $z_{ls}$ and $z_{d}$ depend essentially on pre-recombination physics and are not sensibly affected by late-time parameters.

We complement the analysis by fitting the luminosity distances to JLA \cite{JLA} and Pantheon \cite{Pantheon} type Ia supernovas. JLA permits the supernovas recalibration with the model under test, while in Pantheon the calibration is model independent. As we will see, the two surveys lead to slightly different results. As gaussian priors (see Table I), we will take the Riess {\it et al.} local value of the Hubble-Lema\^itre parameter $h = 0.7348\pm 0.0166$ \cite{Riess18}, and the Cooke {\it et al.} value for the baryonic density parameter $\Omega_{b0} h^2 = 0.02226 \pm 0.00023$, which comes from nucleosynthesis constraints \cite{Cooke}.


Our tests will be performed with two different models. The first is the standard model, for which the indication of a tension has been manifest. This tension will be verified by testing an extension of the standard model given by the generalised Chaplygin gas \cite{gCg1,gCg2,gCg3,gCg4,gCg5,gCg6,gCg7,gCg8}, with a Hubble function given, with the addition of radiation, by
\begin{equation}
E(z)^2 =  \left[ (1 - \Omega_{m0}) + 
\Omega_{m0} (1 + z)^{3(1 + \alpha)} \right]^{1/(1 + \alpha)} + \Omega_{R0}\, (1 + z)^4.
\end{equation}
The binomial expansion of the brackets has a leading term $\Omega_{m0} (1+z)^3$, which means that, for the present purpose of background tests, the baryonic content can be absorbed in the above defined gas. For $\alpha = 0$ we recover the standard $\Lambda$CDM model. Perturbative tests are outside the scope of this paper, but let us comment that, although the adiabatic gCg is ruled out by the observed matter power spectrum owing to oscillations and instabilities \cite{waga}, some non-adiabatic versions have zero sound speed and present a good concordance when tested against background and LSS observations \cite{wands21,wands22,wands23,non-adiabatic1,non-adiabatic2,wang,cassio,pedro,micol}. 

In Fig. 2 (left panel) we show the Hubble parameter $E(z)$ at $z = 17.2$ as function of $\alpha$ for $\Omega_{m0} = 0.31$. It evidences a suppression for negative values of $\alpha$, which leads through (\ref{1})-(\ref{2}) to more negative values of $T_{21}$. In the right panel we plot this temperature as a function of $\alpha$. For positive values of  $\alpha$, the temperature does not change significantly, but it suffers a strong variation in the range $-1 < \alpha < 0$ (for $\alpha < -1$ the gCg inverts its behaviour, acting as a cosmological constant at early times and as matter in the asymptotic future).

\begin{figure*}
 \begin{center}
 \includegraphics[width=0.42\textwidth]{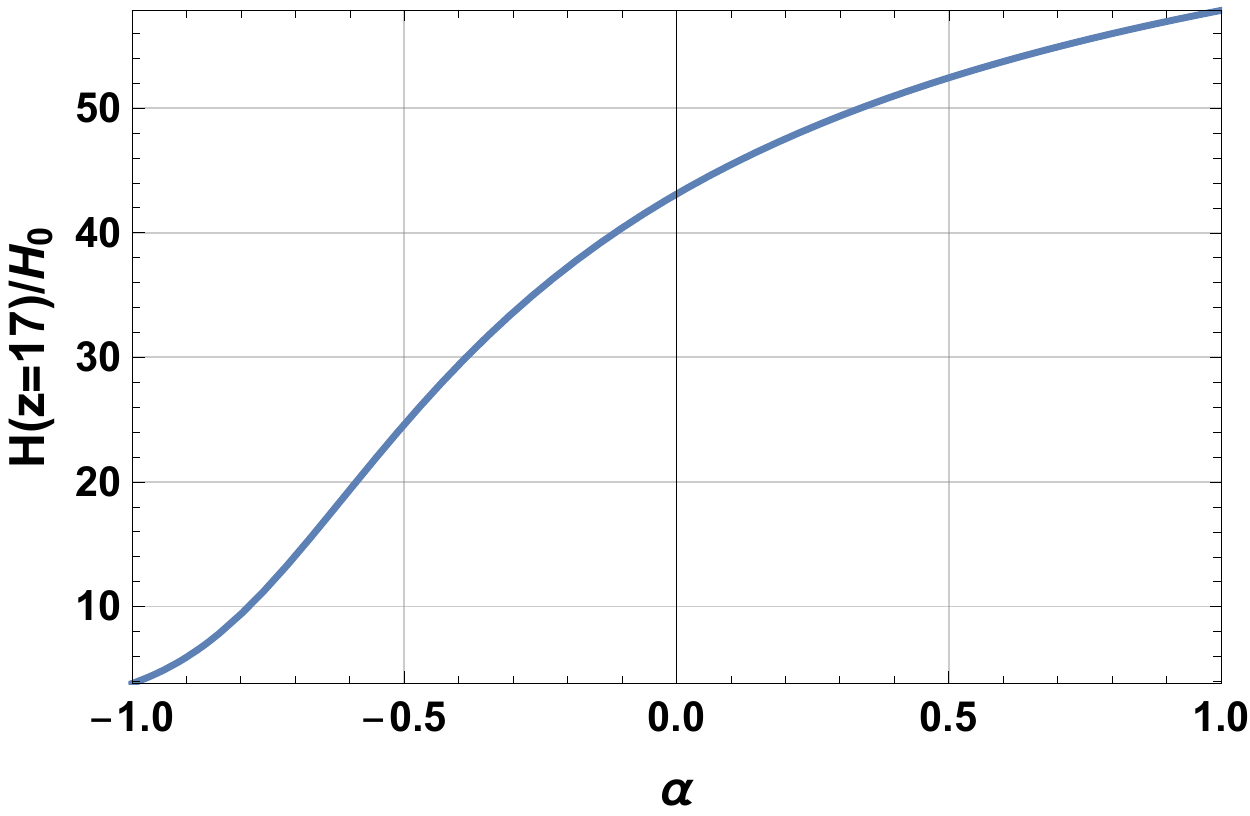}  \quad \quad\includegraphics[width=0.42\textwidth]{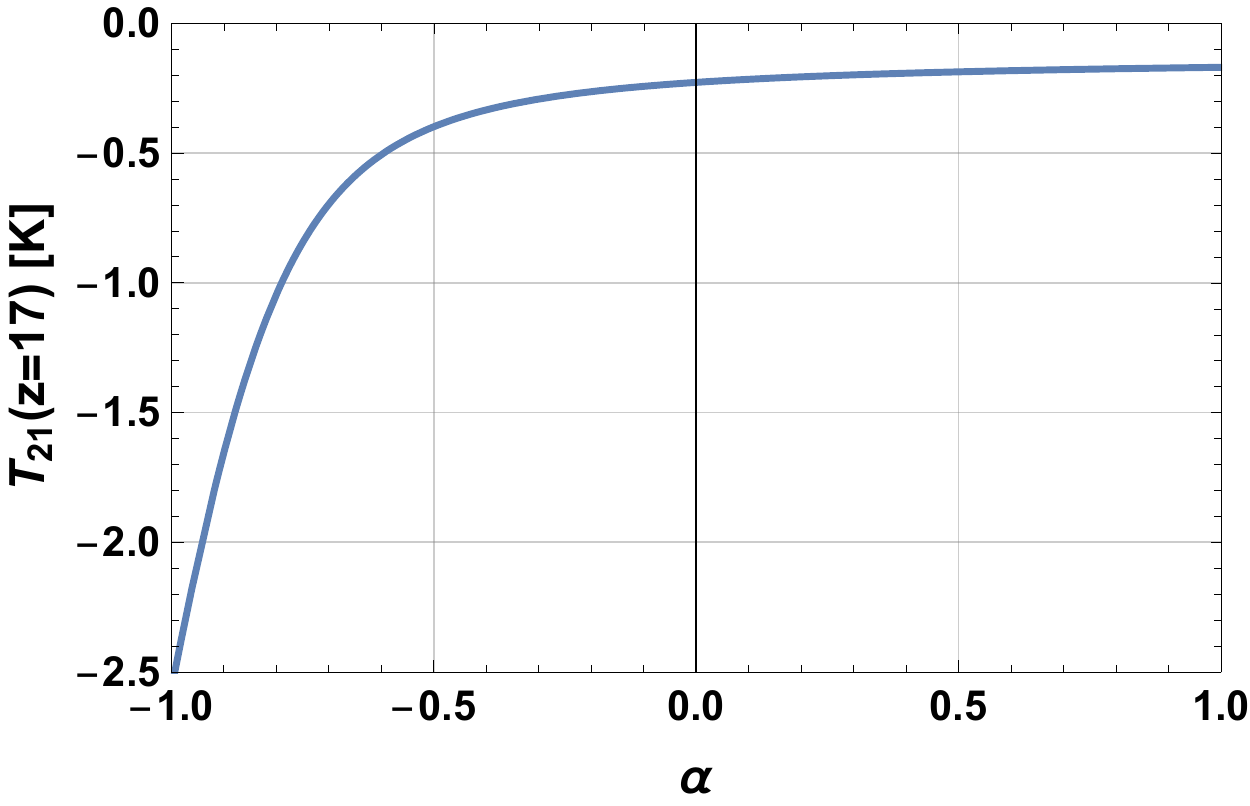}
  \end{center}
  \vspace{-0.3in}
  \caption{$E(z=17.2)$ (left) and $T_{21}(z=17.2)$ (right) as functions of $\alpha$.}
 \label{bxm}
 \end{figure*}

The $1\sigma$ and $2\sigma$ confidence regions for EDGES and JLA observations are shown in the left panel of Fig. 3 and the corresponding likelihoods for $\alpha$ can be seen in the right panel. There is a clear preference for negative values of the interaction parameter, i.e. an energy flux from the dark energy to the dark matter components of the generalised Chaplygin gas. This preference is also clear in the combined likelihoods shown in Fig. 4. In Fig. 5 we present the likelihoods and confidence regions of the joint analysis including $\theta_*$ and $r_d^h$, which also includes the priors imposed on the Hubble-Lema\^itre parameter and baryons density. 
The combined data can be well accommodated with a matter density parameter $\Omega_{m0} \approx 0.3$ and a Hubble-Lema\^itre parameter $h \approx 0.7$, with a best-fit for the gCg parameter slightly negative. The standard model with $\alpha = 0$ is not ruled out, but its concordance value for the matter density parameter is in slight tension with Planck's best-fit \cite{Planck}. This tension is indeed manifest if only the EDGES data are considered. The $2\sigma$ confidence intervals are given on Table II.

Some mechanisms have been recently proposed in order to explain the anomaly observed in the EDGES measurement of the 21 cm line at high redshift. The decay of dark matter into particles of the Standard Model is one of them \cite{barkana,decay}. Interaction between dark matter and dark energy is another possibility. The Chaplygin gas model studied here fits in the spirit of the last proposal.
Such model has already shown many interesting results even at perturbative level if non-adiabatic perturbations are allowed. In what concerns the EDGES data, the results obtained are compatible with SNe Ia data, and the $\Lambda$CDM particular case is admitted only marginally. Such tension is alleviated if CMB and BAO data are included, but the $\Lambda$CDM model is not the preferred scenario. 

We have limited our analyses to background tests, because the main effect into play is the attenuation of the Hubble function at intermediate redshifts in interacting models as compared to the standard model. This alleviates the tension in the EDGES signal, as seen in Fig. 2. Including in the analysis the position of the CMB first peak ($\theta_*$) and the acoustic scale $r_d^h$ shows that a combined test with other observables gives an interaction parameter around zero, in tension with the 21cm line alone (or combined with supernovas). The inclusion of the full CMB spectrum would not change this result, as it also favours $\alpha \approx 0$, as shown e.g. in \cite{micol}.

\section*{Acknowledgements}

We are thankful to J.A. de Freitas Pacheco for useful discussions. Work partially supported by CNPq and FAPES.



\begin{figure*}
 \begin{center}
 \includegraphics[width=0.45\textwidth]{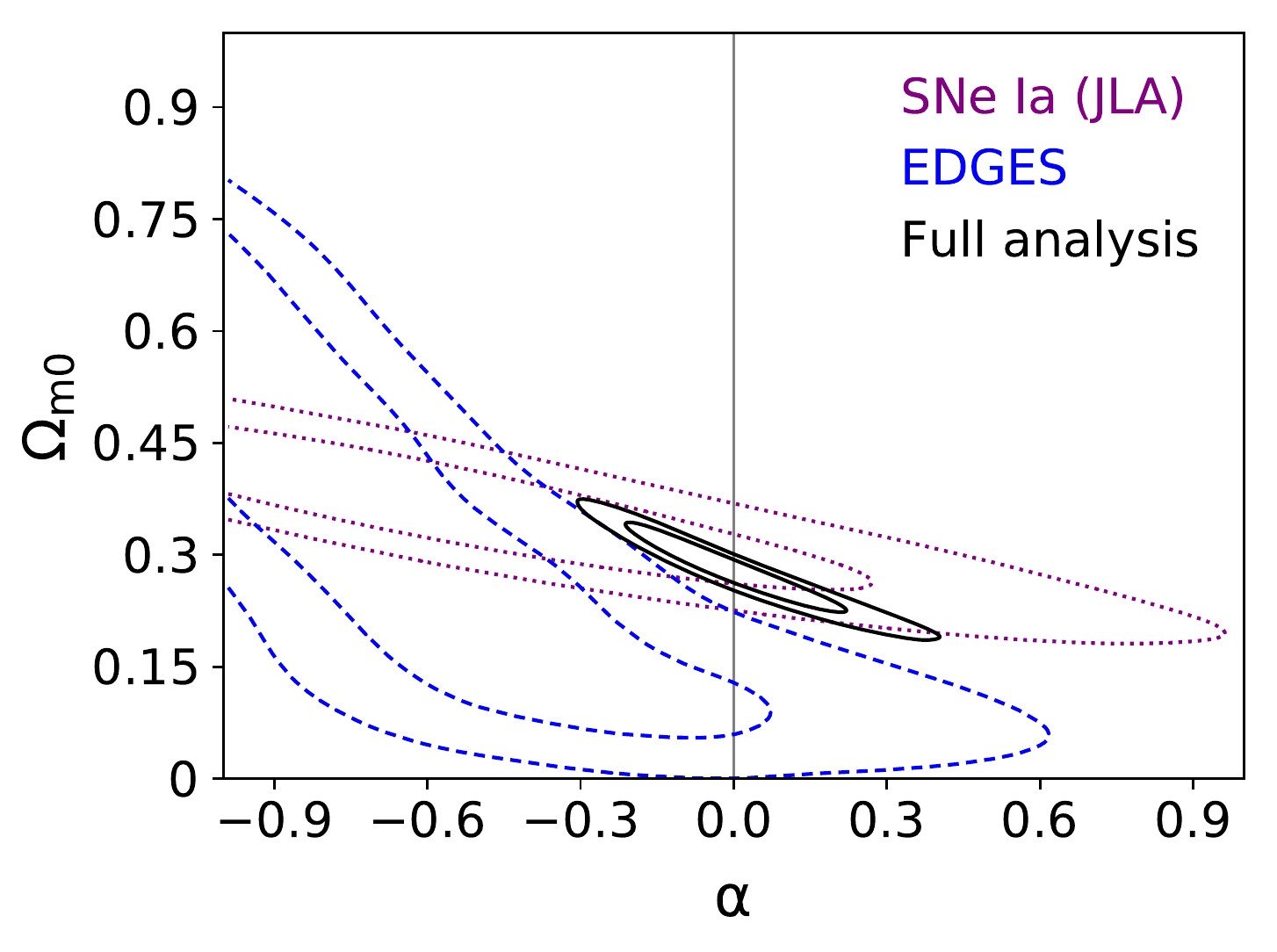} \quad \quad  \includegraphics[width=0.39\textwidth]{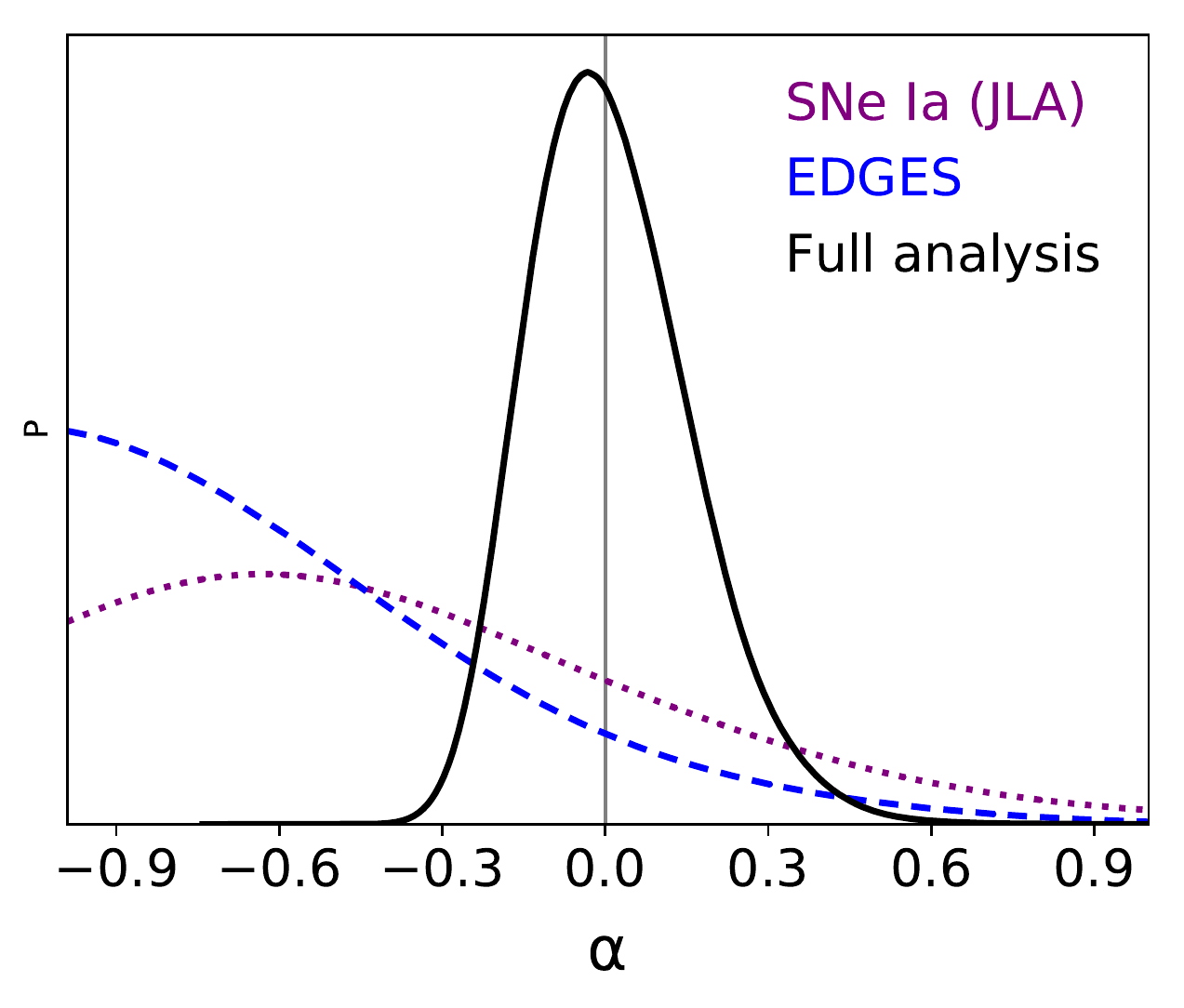}  
  \end{center}
  \vspace{-0.3in}
  \caption{Left panel: $\Omega_{m0} \times \alpha$ confidence regions for the $21$-cm line only (blue), JLA SNe Ia only (purple) and full analysis including $\theta_*$, $r_d^h$ and priors  (black). Right panel: $\alpha$ normalised probability density functions. The priors adopted are presented on Table \ref{tab:priors}.}
 \label{bxm}
 \end{figure*}

\begin{figure*}
 \begin{center}
 \includegraphics[width=0.7\textwidth]{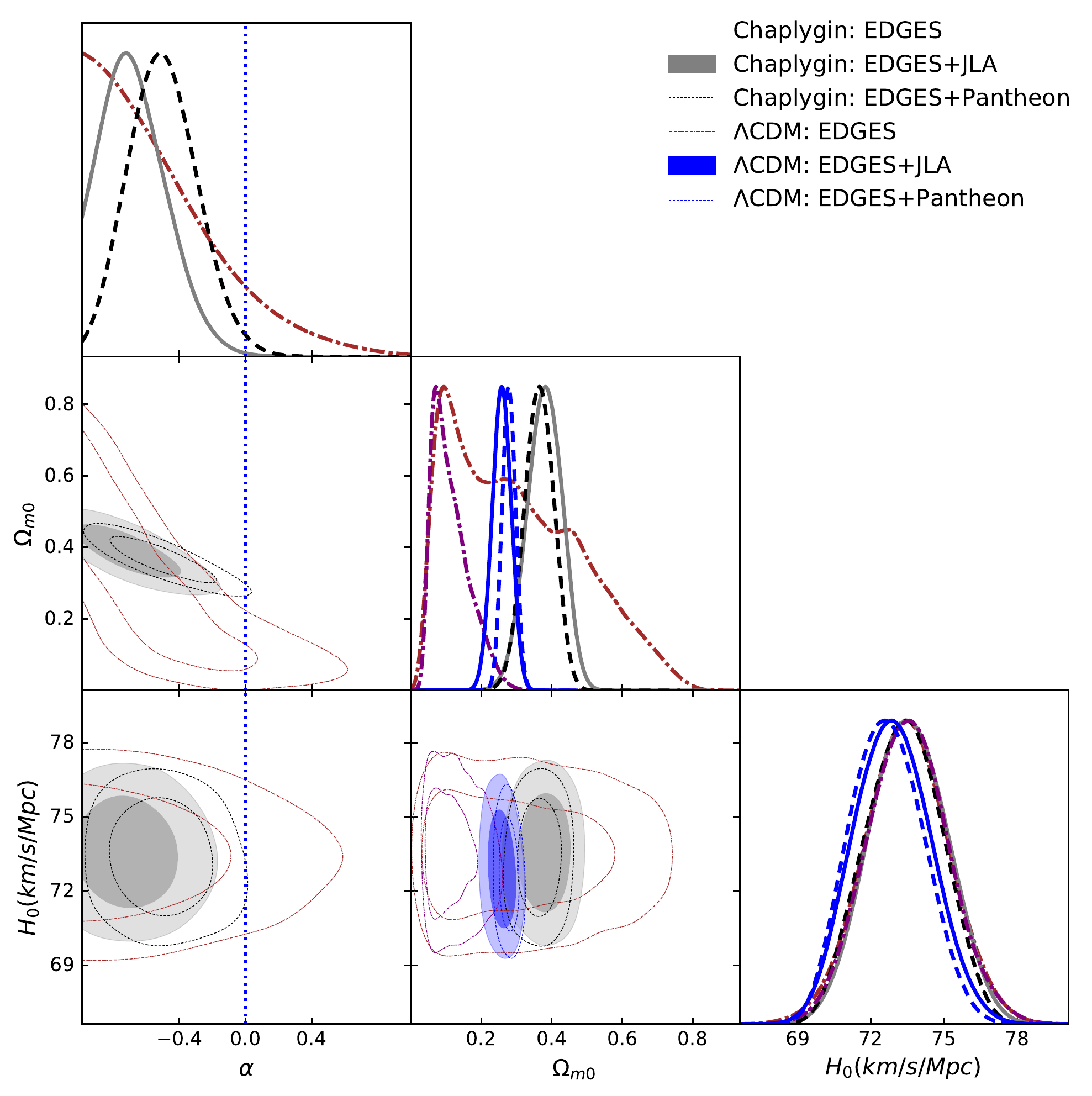} 
  \end{center}
  \vspace{-0.3in}
  \caption{Probability distributions and confidence regions for $\Omega_{m0}$, $H_0$ and $\alpha$ obtained with 21-cm line only and its combinations with SNe Ia data.}
 \label{bxm2}
 \end{figure*}

\begin{figure*}
 \begin{center}
 \includegraphics[width=0.7\textwidth]{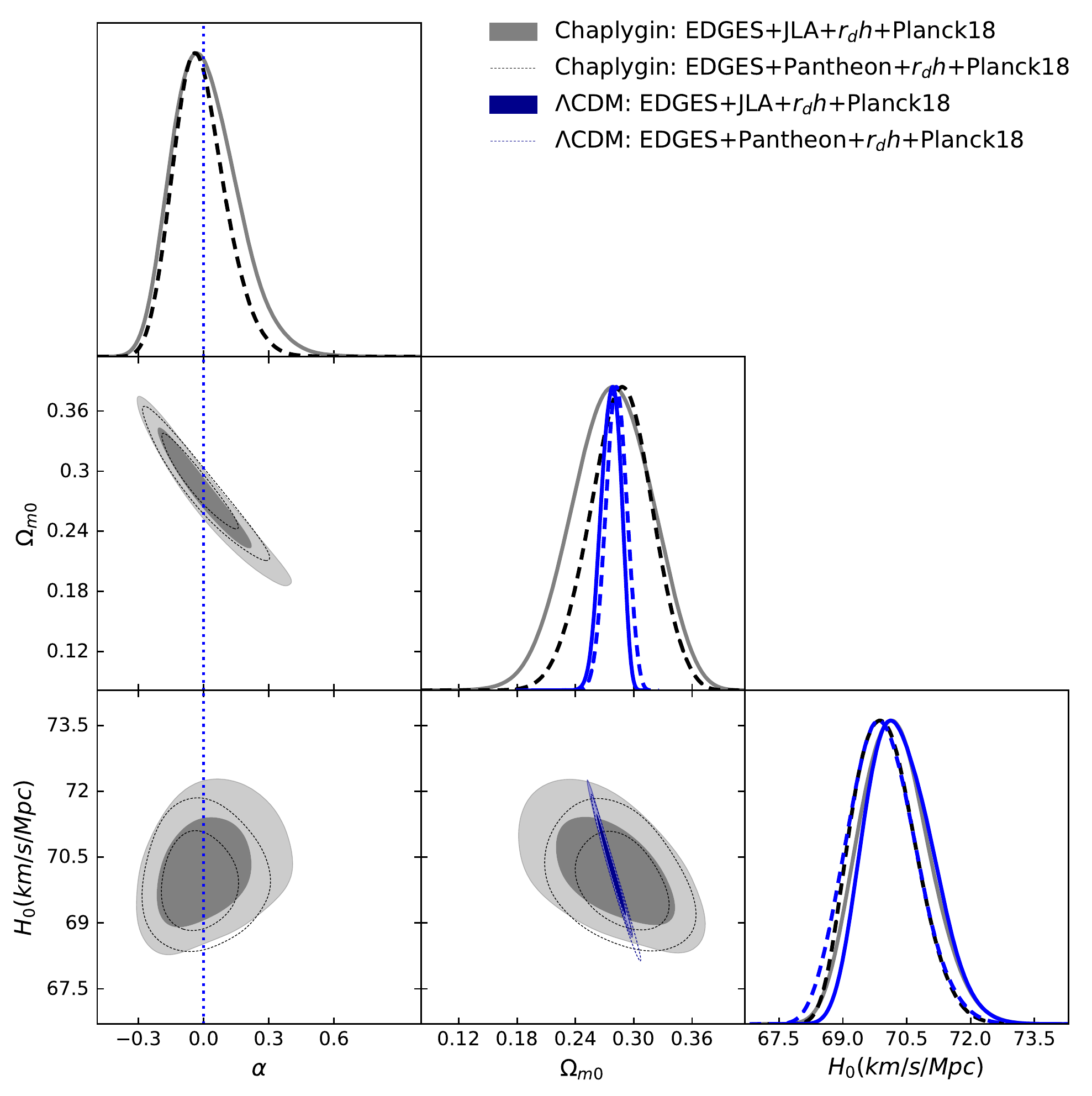} 
  \end{center}
  \vspace{-0.3in}
  \caption{Joint analysis probability distributions and confidence regions for $\Omega_{m0}$, $H_0$ and $\alpha$.}
 \label{bxm2}
 \end{figure*}

\begin{table}
\begin{tabular}{l|c c }
\hline
Free parameters & Priors \ & \\
\hline
$H_0$ (km/s/Mpc) \ & $73.48\pm1.66$ & Gaussian \cite{Riess18} \\
$\Omega_{b0}h^2$ & \ $0.02226\pm0.00023$  \ & Gaussian \cite{Cooke} \\
$\Omega_{dm0}h^2$ & $[  0.001, 1 ]$ & Uniform\\
$\alpha$ & $[  -0.99, 1 ]$ & Uniform\\
\hline
$\alpha_{SNe}$ & $[ 0, 1 ]$ & Uniform\\
$\beta_{SNe}$ & $[ 0, 4 ]$ & Uniform\\
$M$ & $[ -22,-16 ]$ & Uniform\\
$\Delta_M$ & $[ -1,1 ]$ & Uniform\\

\hline
\end{tabular}
\\
\caption{Free parameters and their adopted priors. The parameter space was explored through PyMultiNest module \cite{pymultinest}, with 1500 livepoints and sampling efficiency equal to 0.5. All other PyMultiNest parameters were set to default ones.} \label{tab:priors}
\end{table}

\begin{table}
\begin{tabular}{lc|c c | c}
\hline
&Data                       & $\alpha$                & $\Omega_{m0}$         & $\chi^2_r$                   \\
\hline
\multirow{5}{0 cm} 
              &EDGES+priors                          & $ < 0.24$               & $0.30^{+0.34}_{-0.27}$  & $10^{-9}$ \\
              &EDGES+JLA+priors                      & $ < -0.31$              & $0.38^{+0.09}_{-0.10} $  &   $0.931$ \\
              &EDGES+JLA+$r_d^h$+Planck18+priors      & $0.01^{+0.30}_{-0.27}$   & $0.28\pm0.08$       &   $0.937$  \\
              &EDGES+Pantheon+priors                 & \ $-0.50\pm0.40$ \             & \ $0.36\pm0.08$        &   \ $0.987$ \ \\
              &EDGES+Pantheon+$r_d^h$+Planck18+priors & $-0.02^{+0.24}_{-0.22}$ & $0.29\pm0.06$          & $0.987$  \\ \hline
\end{tabular}
\\
\caption{$2\sigma$ intervals with JLA and Pantheon and its reduced $\chi^2$.}
\end{table}



\end{document}